\pdfoutput=1
\documentclass[a4paper,nobibnotes,twocolumn,prl,amsmath,amssymb,%showpacs,
superscriptaddress]{revtex4-1}
\usepackage{graphicx,epstopdf}
\usepackage{braket}
\usepackage{pdfpages}
\makeatletter
\AtBeginDocument{\let\LS@rot\@undefined}
\makeatother

\newcommand{\identity}{\ensuremath{\openone}}
\newcommand{\tr}{\mathrm{tr}}
\newcommand{\Op}[1]{\ensuremath{\mathsf{\hat{#1}}}}
\newcommand{\Abs}[1]{\left|#1\right|}
\newcommand{\simplex}{\operatorname{splx}}
\newcommand{\PE}{\operatorname{PE}}
\newcommand{\SQ}{\operatorname{SQ}}
\newcommand{\SU}{\operatorname{SU}}
\newcommand{\Hadamard}{\operatorname{H}}
\newcommand{\Phase}{\operatorname{S}}

\newcommand{\pop}{\operatorname{pop}}
\newcommand{\avg}{\operatorname{avg}}
\newcommand{\diag}{\operatorname{diag}}
\newcommand{\epsavg}{\varepsilon_{\avg}}
\newcommand{\microsec}{$\mu$s}
\newcommand{\SigmaX}{\Op{\sigma}_x}
\newcommand{\SigmaY}{\Op{\sigma}_y}
\newcommand{\SigmaZ}{\Op{\sigma}_z}
\newcommand{\SigmaPlus}{\Op{\sigma}_{\!+}}
\newcommand{\SigmaMinus}{\Op{\sigma}_{\!-}}
\newcommand{\Fsm}{F_{\text{sm}}}
\newcommand{\JLI}{J_{\text{LI}}}
\newcommand{\JPE}{J_{\text{PE}}}
\newcommand{\Jsm}{J_{\text{sm}}}
\newcommand{\tildeOp}[1]{\mathsf{\tilde{#1}}}
\newcommand{\Norm}[1]{\left\lVert#1\right\rVert}
\newcommand{\DynMap}{\mathcal{E}}
\newcommand{\unit}[2]{\ensuremath{#1\,\text{#2}}}
\newcommand{\Pout}{\ensuremath{P_{\text{outside}}}}
\newcommand{\Complex}{\mathbb{C}}
\newcommand{\Real}{\mathbb{R}}

\newcommand{\Stanford}{Edward L. Ginzton Laboratory, Stanford University, Stanford, CA 94305, USA.}
\newcommand{\ARL}{U.S. Army Research Laboratory, Computational and Information Sciences Directorate, Adelphi, MD 20783, USA.}
\newcommand{\BQIC}{Berkeley Center for Quantum Information and Computation, Berkeley, California 94720 USA}
\newcommand{\DeptChem}{Department of Chemistry, University of California, Berkeley, California 94720 USA}
\newcommand{\Aarhus}{Department of Physics and Astronomy, Aarhus University, 8000 Aarhus C, Denmark.}
\newcommand{\Kassel}{Theoretische Physik, Universit\"{a}t Kassel, Heinrich-Plett-Str.\ 40, D-34132 Kassel, Germany}

\makeatletter
\def\Fig{\@ifnextchar[{\@FigWith}{\@FigWithout}}
\def\@FigWith[#1]#2{Fig.~\ref{fig:#2}\,(#1)}
\def\@FigWithout#1{Fig.~\ref{fig:#1}}
\def\Figure{\@ifnextchar[{\@FigureWith}{\@FigureWithout}}
\def\@FigureWith[#1]#2{Figure~\ref{fig:#2}\,(#1)}
\def\@FigureWithout#1{Figure~\ref{fig:#1}}
\makeatother

\begin{document}

\title{Charting the circuit QED design landscape using optimal control theory}

\author{Michael H.\ Goerz}
\email{Correspondence: goerz@stanford.edu}
\altaffiliation{Current affiliations: \Stanford, and \ARL}
\affiliation{\Kassel}
\author{Felix Motzoi}
\altaffiliation{Current affiliation: \Aarhus}
\affiliation{\BQIC} \affiliation{\DeptChem}
\author{K.\ Birgitta Whaley}
\affiliation{\BQIC} \affiliation{\DeptChem}
\author{Christiane P.\ Koch}
\affiliation{\Kassel}

\date{\today}

\begin{abstract} % Limit: 150 words, no references (current word count: 145)
  With recent improvements in coherence times, superconducting transmon qubits
  have become a promising platform for quantum computing. They can be flexibly
  engineered over a wide range of parameters, but also require us to identify an
  efficient operating regime. Using state-of-the-art quantum optimal control
  techniques, we exhaustively explore the landscape for creation and removal of
  entanglement over a wide range of design parameters. We identify an
  optimal operating region outside of the usually considered strongly dispersive
  regime, where multiple sources of entanglement interfere simultaneously, which
  we name the quasi-dispersive straddling qutrits (QuaDiSQ) regime.  At a chosen
  point in this region, a universal gate set is realized by applying microwave
  fields for gate durations of \unit{50}{ns}, with errors approaching the limit
  of intrinsic transmon coherence.  Our systematic quantum optimal control
  approach is easily adapted to explore the parameter landscape of other quantum
  technology platforms.
\end{abstract}

\maketitle

For quantum technology to unfold its full potential, ultimate performance bounds
must be known.
This concerns all relevant steps for operating a device, such as state
preparation or quantum gate implementation.
One such bound is the empirical ``quantum speed limit'' that determines the
shortest possible duration to carry out the task at hand~\cite{CanevaPRL09,
GoerzJPB11,SorensenN2016}.
Quantum optimal control (QOC)~\cite{GlaserEPJD2015, KochJPCM2016} has grown into
a versatile tool for identifying these performance bounds.
Typical control tasks include the preparation of nonclassical states, as shown
in an experiment with a Bose-Einstein condensate~\cite{FrankNC2014}, or the
creation of entanglement and quantum error correction, as demonstrated with
diamond spin systems~\cite{DoldeNC2014,WaldherrN2014}.
To date, these tasks have been optimized for known, fixed parameters of the
respective system.
Here, we show that  a fully numerical QOC approach can go even further and,
using the most advanced control techniques,  can map out the entire  parameter
landscape for the physical system at hand.
To this end, we consider the task of realizing the fastest possible universal
set of gates for two superconducting transmon qubits within the constraints of
current lifetimes.

Superconducting transmon qubits~\cite{JKochPRA07} are one of the most promising
architectures for quantum computing today.
The development of circuit QED~\cite{BlaisPRA2007}, a broad platform for quantum
technology, in particular enabled the entanglement of spatially separated
superconducting qubits via a shared transmission line resonator.
The shared resonator can be used to implement two-qubit gates but is generally
required to be decoupled when single-qubit gates are carried out.

Three principal approaches have been used to couple superconducting qubits via a
resonator:
fixed-frequency~\cite{LeekPRB2009,ChowPRL2011,ChowPRL2012,PolettoPRL2012,
ChowNJP2013,ChowNC2014,CorcolesNC2015,EconomouPRB2015,CrossPRA2015,SheldonPRA2016},
tunable frequency~\cite{DiCarloN09,GhoshPRA2013,EggerSST2014}, and tunable
coupling~\cite{WallquistPRB2006,SrinivasanPRL2011,AllmanPRL2014,AndersenPRA2015,McKayPRAP2016}.
Fixed-frequency transmons require the least technological overhead but also make
the realization of gates most difficult.
Adding more overhead in terms of dedicated control lines for the purpose of
qubit driving~\cite{ChowPRL2011} or frequency biasing~\cite{JKochPRA07} can
speed up gate implementation but comes at the cost of additional noise
sources~\cite{MartinisPRL2005, ConstantinPRL2007, JKochPRA07,MotzoiPRA2013}.

In all these approaches, a wide range of possible parameters can be engineered.
However, the parameter regimes where single- and two-quit gates can be
faithfully operated are typically very different and the optimal choice of
parameters is then not obvious.
Identification of the optimal operating parameters is a well-defined control
problem that we address here with QOC\@.
While QOC has been used to realize specific quantum gates on superconducting
qubits~\cite{MotzoiPRL2009, RebentrostPRL2009, ChowPRA2010, EggerSST2014,
EggerPRL2014, CrossPRA2015, ZahedinejadPRL2015,TheisPRA2016},  no systematic
exploration of the full parameter space has been undertaken to date.
In particular, most prior work has focused exclusively on the dispersive regime
regime and explicitly avoided the regime of strong coupling.
We show here that a fully numerical approach combined with advanced QOC
techniques allows us to map the entire parameter landscape without restrictions
due to approximations or model reduction.
We can thus identify the global quantum speed limit for a universal set of gates
for transmon qubits and analyze how gate errors vary with
qubit-cavity-couplings, resonances, and cavity-mediated decay.
This guides the decision for specific working points that promise a successful
implementation of universal quantum computing using superconducting qubits.
In particular, our results show that fast operation of both the single- and
two-qubit gates needed for universal computation can be implemented in parameter
regimes outside that typically explored to date.
The results thus provide critical information for design decisions in circuit
QED, or similarly complex quantum architectures.

\section{Results}

\subsection{Model and Parameters}

Two superconducting transmons with a common transmission line resonator can be
modeled by two anharmonic ladders coupled to a driven harmonic
oscillator~\cite{JKochPRA07}.
In the frame rotating with frequency $\omega_r$ and within the rotating wave
approximation, the Hamiltonian reads
\begin{eqnarray}
  \Op{H}
  &=& \sum_{q=1,2} \hbar \left[ \delta_q \Op{b}^{\dagger}_q \Op{b}_q
    + \frac{\alpha_q}{2} \Op{b}^{\dagger}_q\Op{b}^{\dagger}_q\Op{b}_q\Op{b}_q
    + g \left(\Op{b}^{\dagger}_q \Op{a} + \Op{b}_q \Op{a}^{\dagger} \right)
    \right] \nonumber \\
    &&  + \hbar \delta_c \Op{a}^{\dagger}\Op{a}
    + \frac{\hbar}{2} \left(\epsilon(t) \Op{a} +
                            \epsilon^*(t) \Op{a}^{\dagger}\right)\,,
\label{eq:ham_full}
\end{eqnarray}
with $\delta_{j} = \omega_{j} - \omega_r$ ($j=1,2,c$), where $\omega_c$ is the
resonator (``cavity'') frequency,  $\omega_{1,2}$ is the frequency of the first
(second) qubit, $\alpha_{1,2}$ is the qubit anharmonicity, $g$ is the coupling
strength between each qubit and the resonator, and $\epsilon(t)$ is the
microwave control field in the cavity.
Taking $\epsilon(t) \in \Complex$ is equivalent to controlling the pulse
amplitude and phase independently.
It corresponds to a control field $\epsilon(t) \in \Real$  in the non-rotating
frame whose driving frequency may deviate from $\omega_r$.
In the following, we set $\hbar = 1$.

Typically, the Hamiltonian~\eqref{eq:ham_full} is treated perturbatively, in
order to derive an effective model in which the cavity can be integrated out.
This approach is only valid in the ``dispersive regime'' where the qubit-cavity
separation is much larger than the qubit-cavity coupling, $\Abs{\omega_c -
\omega_{1,2}} \gg g$.
This limits the effective interaction $\propto g/(\omega_c - \omega_{1,2})$
between both qubits, except when resonances with higher transmon levels can be
exploited~\cite{PolettoPRL2012,ChowNJP2013}.
Here, we forgo such a treatment in lieu of solving Eq.~\eqref{eq:ham_full}
numerically, allowing us to explore parameter regimes beyond the dispersive
limit.

In order to limit the number of parameters that have to be varied, we focus on
the two  parameters that capture the essential physics of quantum gate
implementation.
The departure from the dispersive regime is characterized by  $\Delta_c / g$,
with $\Delta_c = \omega_c - \omega_1$.
Secondly, resonances of the form $\Abs{\omega_1 - \omega_2} \approx n
\alpha_{1,2}$, $n=1,2$, between different levels of the two transmons are known
to aid in the implementation of entangling gates~\cite{PolettoPRL2012,
ChowNJP2013}; therefore, it is natural to express the qubit-qubit-detuning
$\Delta_2 = \omega_2 - \omega_1$ in units of $\alpha = \Abs{\alpha_1 +
\alpha_2}/2$.
We can thus explore the entire parameter landscape in terms of $\Delta_c/g$ and
$\Delta_2/\alpha$ by keeping $\omega_1$, $g$, $\alpha_1$, and $\alpha_2$ fixed
and varying $\omega_2$ and $\omega_c$.
At each parameter point $(\Delta_c/g, \Delta_2/\alpha)$, there is a unique
minimum duration (a quantum speed limit) for any gate operation and
pre-specified error.

\begin{table}[t]
  \centering
  \begin{tabular}{lcl}
    $\omega_1/2\pi$ &  $=$      & $  \unit{6.0}{GHz}$                                                            \\
    $\omega_2/2\pi$ &  $=$      & $  5.0$ -- $\unit{7.0}{GHz}$ (vary); $\Delta_2 \equiv \omega_2 - \omega_1$     \\
    $\omega_c/2\pi$ &  $=$      & $  4.5$ -- $\unit{9.0}{GHz}$ (vary); $\Delta_c \equiv \omega_c - \omega_1$     \\
    $\begin{matrix}
      \alpha_1/2\pi \\
      \alpha_2/2\pi
    \end{matrix} $  &
                      $\begin{matrix}
                        =\\
                        =
                        \end{matrix} $  &
                                    $\left.\begin{matrix}
                                      \!\unit{-290}{MHz}\\
                                      \!\unit{-310}{MHz}
                                    \end{matrix}\right\}$
                        $\alpha \equiv \frac{1}{2}\Abs{\alpha_1 + \alpha_2}$  \\
    $g       /2\pi$ &  $=$      & $\unit{   70}{MHz}$                                                            \\
    $\gamma  /2\pi$ &  $=$      & $\unit{0.012}{MHz}$;                $\tau_{\gamma} = \unit{13.3}{\microsec}$~\cite{CrossPRA2015}    \\
    $\kappa  /2\pi$ &  $=$      & $\unit{ 0.05}{MHz}$;  \hspace{1.5mm}$\tau_{\kappa} = \unit{3.2}{\microsec}$~\cite{PetererPRL2015}
  \end{tabular}
  \caption{System parameters}
\label{tab:parameters}
\end{table}

All parameters are listed in Table~\ref{tab:parameters}.
We include in our model the most relevant source of dissipation, spontaneous
decay of the qubits with a decay rate $\gamma$ (and the associated qubit
lifetime $\tau_{\gamma}$), and spontaneous decay of the cavity with decay rate
$\kappa$ (lifetime $\tau_{\kappa}$). We assume the same decay rate for both
qubits for convenience only.
Different rates would result in correspondingly different error limits for
single-qubit operations on each qubit, but not affect results otherwise.

The full equation of motion is the Liouville-von-Neumann equation with the
Hamiltonian~\eqref{eq:ham_full}, and Lindblad operators $\Op{a}$, $\Op{b}_1$,
and $\Op{b}_2$.
We encode the logical subspace, denoted by $\ket{00}$, $\ket{01}$, $\ket{10}$,
$\ket{11}$, as those ``dressed'' eigenstates of the Hamiltonian (for
$\epsilon(t) = 0$) that have the largest overlap with the ``bare'' states
$\ket{ijn} = \ket{000}$, $\ket{010}$, $\ket{100}$, $\ket{110}$, where $i$ $j$,
and $n$ are the quantum numbers for the first transmon, second transmon, and the
cavity, respectively.

\subsection{Qubit interaction, entanglement, and spontaneous decay losses for
zero external drive}

\begin{figure*}[tb]
  \centering
  \includegraphics{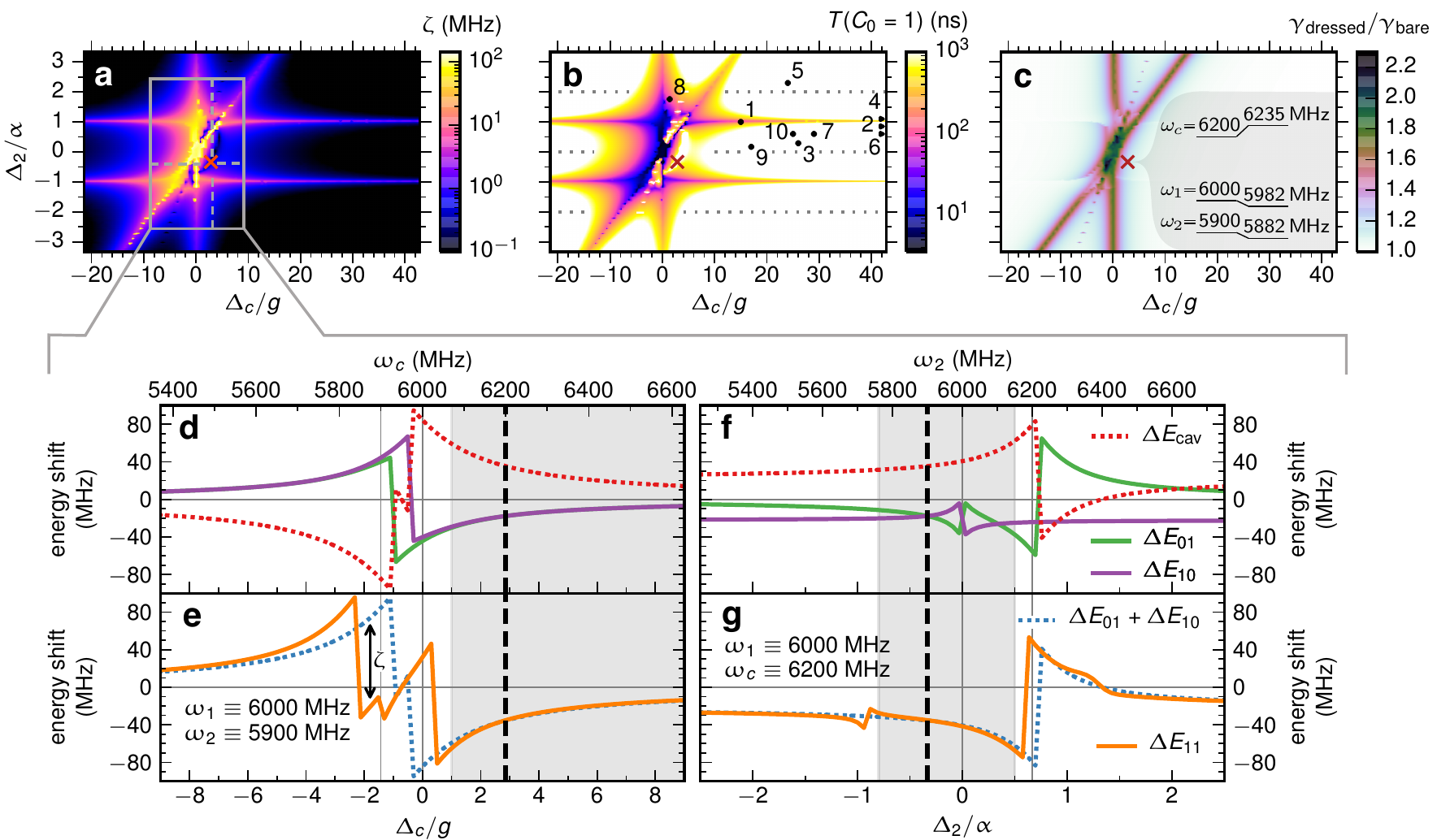}
  \caption{Field-free properties of the parameter landscape.
  (a) Always-on interaction energy $\zeta$ resulting from the cavity-induced
  shift of the (dressed) qubit levels, Eq.~\eqref{eq:zeta}.
  (b) Gate duration after which the field-free evolution produces a fully
  entangling gate. The points labeled 1--10 mark the parameters for some
  existing implementations of entangling gates for fixed-frequency
  transmons, cf.~Refs.~\cite{LeekPRB2009,ChowPRL2011,ChowPRL2012,
  PolettoPRL2012,ChowNJP2013,ChowNC2014,CorcolesNC2015,EconomouPRB2015,
  CrossPRA2015,SheldonPRA2016}. For points 4,~2,~6, $\Delta_c/g$ takes a value
  outside of plotted region (58, 95, and 43, respectively). The gate durations
  are  150$^{(1)}$, 220$^{(2)}$, 110$^{(3)}$, 200$^{(4)}$, 500$^{(5)}$,
  350$^{(6)}$, 350$^{(7)}$, 50$^{(8)}$, 120$^{(9)}$, and
  $\unit{200^{(10)}}{ns}$.  The horizontal gray dotted lines indicate
  $\Delta_2/\alpha = -2, 0, 2$, for visual reference.
  (c) Ratio of the dressed qubit decay rate to the bare qubit decay rate. The
  point labeled by the red X indicates possible parameters where to implement
  a full universal set of gates. The bare qubit and cavity frequencies
  $\omega_1, \omega_2, \omega_c$ at this point are shown in the inset, together
  with their ``dressed'' value $E_{10}$, $E_{01}$, $E_{001}$, i.e., the
  eigenenergies of the corresponding logical levels, respectively
  the eigenenergy $E_{001}$ of the eigenstate closest to the bare state
  $\ket{001}$.
  (d,~e) For a horizontal slice through the parameter space as indicated in
  panel a, value of $\Delta E_{01} \equiv E_{01} - \omega_2$,
  $\Delta E_{10} \equiv E_{10} - \omega_1$,
  $\Delta E_{11} \equiv E_{11} - \omega_1 - \omega_2$, and
  $\Delta E_{\text{cav}} \equiv E_{001} - \omega_c$.
  (f,~g) Energy shifts for vertical slice as indicated in panel a.
  In panels (e,~g), the difference between the solid orange and dotted blue
  curve is $\zeta$, see Eq.~\eqref{eq:zeta}.
  The gray region highlights the quasi-dispersive straddling qutrits (QuaDiSQ)
  regime, cf.~\protect\Fig[n]{octmap}. The vertical thick dashed line and thin
  solid lines mark the parameters of point X, and the cavity-qubit or
  qubit-cavity resonances, respectively.
  \vspace{1cm}
}
\label{fig:field-free}
\end{figure*}

Creation of entanglement is typically considered the most difficult task in
implementation of a universal set of gates.
The use of dressed logical states results in a static qubit-qubit interaction
$\zeta$ (for $\epsilon(t) = 0$) that may be exploited to this end.
The interaction is the result of the eigenenergies $E_{00}$, $E_{01}$, $E_{10}$,
and $E_{11}$ being shifted relative to the bare frame, which leads
to~\cite{GoerzJPB11}
\begin{equation}
  \zeta = E_{00} - E_{01} - E_{10} + E_{11}\,.
  \label{eq:zeta}
\end{equation}
The value of $\zeta$ is shown in \Fig[a]{field-free} as a function of the two
parameters $\Delta_c/g$ and $\Delta_2/\alpha$.
For a horizontal and vertical slice through the parameter landscape at the point
marked as X, \Fig[d--g]{field-free} shows how the dressed energy levels vary
with $\Delta_c$ and $\Delta_2$, and how they combine to $\zeta$.
The energy shifts become especially large when any of the (bare) qubits are
near-resonant with the cavity (vertical line in panels a--c at
$\Delta_c=\omega_c-\omega_1=0$), or with the ``anharmonic transition''
(horizontal lines at $\Delta_2 \approx \pm \alpha$). They jump in sign when
crossing through the resonances, see \Fig[d,f]{field-free}.
The diagonal resonance line in panels a--c is for $\omega_c = \omega_2$.
The interaction strength can reach values \unit{>100}{MHz} when the qubit and
the cavity frequencies are very close.
Such a large static interaction would allow for very fast entangling gates.
The field-free evolution for a duration $T$ induces an entangling gate with a
concurrence~\cite{GoerzJPB11}
\begin{equation}
  C_0(\zeta, T) = \Abs{\sin\frac{\zeta T}{2}}\,.
  \label{eq:C_0}
\end{equation}
A perfectly entangling gate is first reached after $T_{\pi}=\pi/\zeta$.
This time is shown in \Fig[b]{field-free} as a function of $\Delta_c/g$ and
$\Delta_2/\alpha$.

The spontaneous decay of the qubit with a decay rate of $\gamma$ implies a lower
bound on the error of any two-qubit gate.
As shown in Methods, for a fixed gate duration $T$, we find this bound to be
\begin{equation}
  \epsavg^{0}(\gamma, T) \approx \frac{8}{10} \gamma T\,.
  \label{eq:eps_0}
\end{equation}
The decay rate of the logical eigenstates may increase relative to the bare
states, due to overlap with excited bare transmon or resonator levels with
shorter lifetimes.
The ratio of the effective (``dressed'') qubit decay rate to the bare decay rate
is shown in \Fig[c]{field-free} as a function of $\Delta_c/g$ and
$\Delta_2/\alpha$.

At the resonance lines, the increase in dissipation is most severe.
However, even in the worst case, it is within only a factor of $\approx 2.3$ of
the bare rate.
This is in contrast to the static interaction $\zeta$ and the duration to
implement an entangling gate, $T_\pi$, whose values span several orders of
magnitude.
Choosing a specific point for gate implementation within the parameter space may
therefore dramatically increase implementation speed without incurring
substantially larger losses.
It remains to be seen, however, whether parameter space points with a large
static interaction and a small $T_\pi$ also allow for implementation of local
operations.

\subsection{Entanglement creation and removal}

\begin{figure*}[th]
  \centering
  \includegraphics{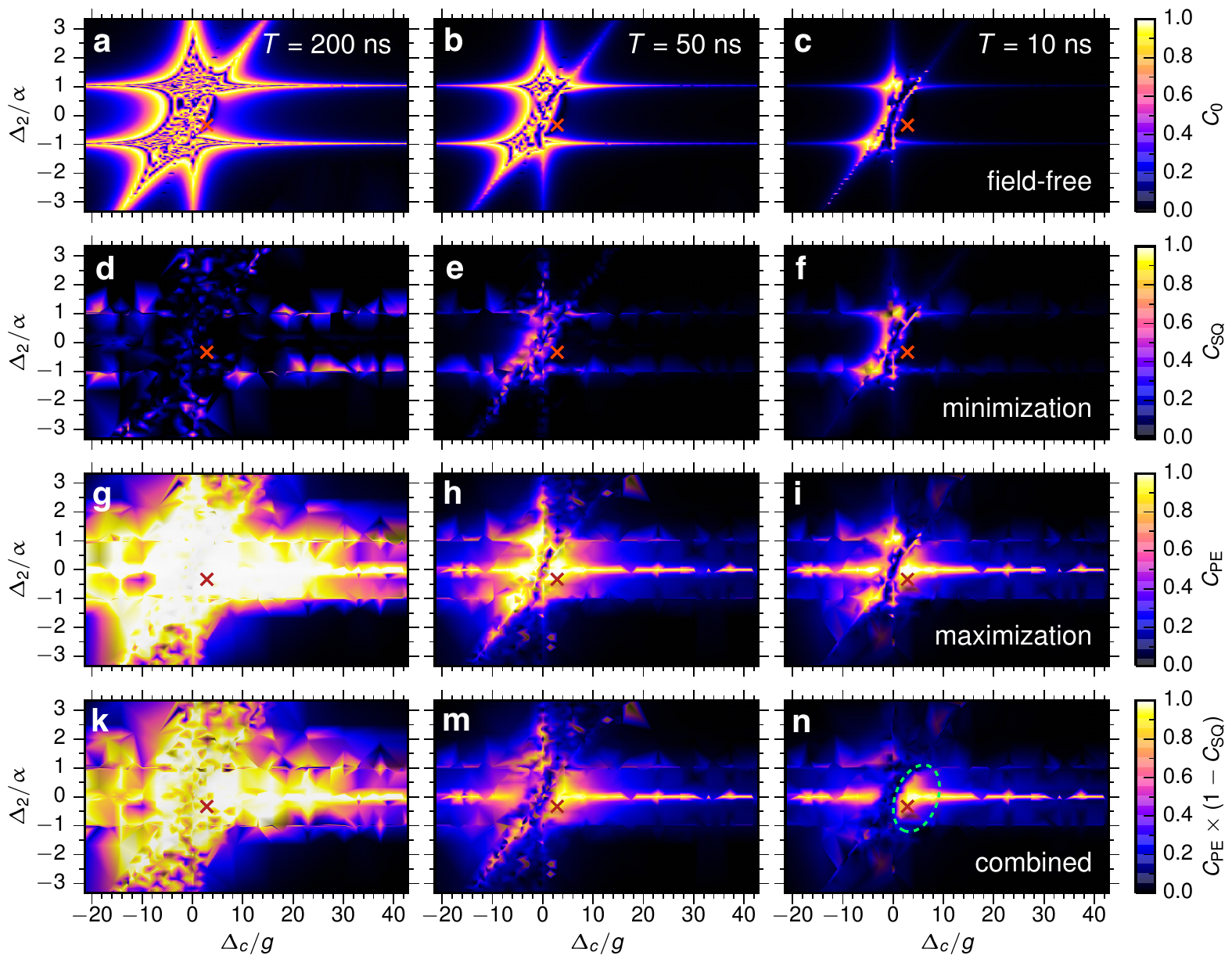}
  \caption{Maximization and minimization of entanglement for varying gate
    duration. (a--c) Concurrence $C_0$ induced by the static interaction energy
    $\zeta$, see~\protect\Fig[a]{field-free}, after field-free evolution for
    \unit{T=200, 50, 10}{ns}. (d--f) Concurrence $C_{\SQ}$ under an
    optimized microwave field that \emph{minimizes} entanglement. (g--i)
    Concurrence $C_{\PE}$ obtained from \emph{maximization} of
    entanglement.  (k--n) Combined measure of success $C_{\PE} \times
    (1-C_{\SQ})$ for the ability to both produce local gates ($C_{\SQ}
    = 0$) and perfectly entangling gates ($C_{\PE} = 1$). In all panels,
    loss from the logical subspace (through excitation or dissipation) is
    indicated as transparency against the (black) background.  The turquoise
    ellipse encircles the ``QuaDiSQ'' region which we identify as optimal for
    the control of entanglement, see text for details. The point marked by the
    red X is a candidate for the implementation of a universal set of gates.
    \vspace{1cm}
    }
\label{fig:octmap}
\end{figure*}

We now consider the use of a control field $\epsilon(t) \neq 0$ for the
realization of quantum gates.
Before targeting specific gates that build up a universal set, we study a
prerequisite---the basic capability to entangle and disentangle.
To this end, we minimize and maximize entanglement, measured by the concurrence,
while minimizing loss due to decay or leakage from the logical subspace, at all
points of the parameter landscape.
At each point, we employ a three-stage optimization, as described in Methods.

Any entanglement that is created or removed by application of a microwave pulse
is relative to the entanglement for the field-free evolution.
Therefore, \Fig[a--c]{octmap} shows the field-free entanglement $C_0(\zeta, T)$
(Eq.~\eqref{eq:C_0})   as a function of $\Delta_c/g$ and $\Delta_2/\alpha$ for
gate durations $T=$200, 50, and \unit{10}{ns}.
The oscillatory behavior  in \Fig[a,~b]{octmap} results from the fact that $T\gg
\pi/\zeta$ for these longer gate times.

The basic capability to carry out local operations, quantified as $C_{\SQ}$, is
analyzed in \Fig[d--f]{octmap}.
Entanglement can be reduced below the field-free values in a large part of the
parameter space by applying suitably shaped pulses, as reflected by the dark
areas in \Fig[d--f]{octmap}.
For long gate durations (\unit{T = 200}{ns}), the concurrence can be brought
close to zero over nearly the entire parameter landscape, see~\Fig[d]{octmap}.
As the pulse duration gets shorter, the parameter region with significant
field-free entanglement becomes smaller (\Fig[b,~c]{octmap}). At the same time,
bringing the concurrence to zero becomes more difficult.
This is true in particular along the resonance lines $\omega_c \approx \omega_1,
\omega_2$, for $-\alpha < \Delta_2 < \alpha$, which is the region for which
there is substantial field-free entanglement even for very short gate durations
(\unit{T = 10}{ns}, \Fig[c]{octmap}). Application of a pulse cannot further
reduce the concurrence, see~\Fig[f]{octmap}.
Thus, a speed limit for pulse-induced removal of the non-local nature of the
interaction, ($C_0 > 0 \rightarrow C_{\SQ} = 0$) is found around \unit{10}{ns}.

The basic capability to \emph{create} entanglement, quantified as $C_{\PE}$,
with microwave control is shown in \Fig[g--i]{octmap}.
Comparison with the field-free entanglement  $C_0$ in \Fig[a--c]{octmap} reveals
that adding microwave controls enlarges the region of parameter space where
entanglement can be created, as expected.
For long gate durations (\unit{T = 200}{ns}), entanglement can be generated in
large parts of the parameter space, in particular in the region $-\alpha <
\Delta_2 < \alpha$ around the resonance of both qubits ($\Delta_2 \approx 0$).
As the gate duration becomes shorter ($T = 50, \unit{10}{ns}$), the region where
perfect entanglers can be implemented shrinks (\Fig[h,~i]{octmap}).  Still, even
for very short gate durations, it is possible to generate pulse-induced perfect
entanglers along the line $\Delta_2 = 0$.

\subsection{Optimal QuaDiSQ regime for implementing a universal set of gates}

The realization of a full universal set of gates~\cite{NielsenChuang} requires a
region in the parameter space that allows for both entangling and local gates.
This is true both for tunable and fixed-frequency qubits, since in the former
case the tuning range should be kept small in order to avoid dephasing (flux)
noise.
To identify such regions, we inspect the product $C_{\PE} \times (1-C_{\SQ})$ in
\Fig[k--n]{octmap}.

It is noteworthy that points with large field-free entanglement $\zeta$,
cf.~\Fig[a]{field-free}, are \emph{not} ideal candidates for fixed-frequency
qubits.
This is because each qubit transition is strongly dressed by the other qubit in
this case.
While it allows for the easy realization of an entangling (CPHASE) gate in under
\unit{10}{ns}~\cite{ChenPRL2014}, cf.~\Fig[b]{field-free}, it prevents
single-qubit gate operations independent of the state of the other qubit.
Similarly, the horizontal line in \Fig[m-n]{octmap} suffers from a resonance
between the qubits ($\Delta_2 = 0$) which impairs their individual
addressability despite the absence of field-free entanglement.

We conclude that an optimal regime for  fixed-frequency transmons requires
simultaneously (1) very small static entangling strength $\zeta$ and (2) a small
dispersive parameter $(\omega_{1,2} - \omega_c)/g$, while (3) also avoiding
resonance of the qubit frequencies.
This is the case in the region encircled by the turquoise dashed ellipse in
\Fig[n]{octmap}, for which we coin the name \emph{Quasi-Dispersive Straddling
Qutrits} (QuaDiSQ).

The quasi-dispersive regime of cavity-QED is defined as $1 < (\omega_{1,2} -
\omega_c)/g < 10$, between the near-resonant ($(\omega_{1,2} - \omega_c)/g < 1$)
and the dispersive ($(\omega_{1,2} - \omega_c)/g > 10$)
regime~\cite{SchusterN2007}, both of which are excluded by conditions (1)--(3).
Meanwhile, the minimal static coupling $\zeta$ between the qubits (cf.\
black/white disc in \Fig[a,b]{field-free}, and crossing solid orange and dotted
blue lines in \Fig[e,f]{field-free}) is enabled by one transmon frequency being
situated between the first and second ($\Ket{1} \rightarrow \Ket{2}$) transition
of the other transmon (hence ``qutrits''), $-1 < \Delta_2/\alpha <1$.
This region is reminiscent of the so-called straddling regime of single-transmon
circuit QED~\cite{JKochPRA07}, with the distinction that the qutrits are
straddling one another here, rather than the cavity.
This allows the level repulsion to act with opposite signs, see
\Fig[d,f]{field-free}, effectively resulting in destructive interference and
avoiding unwanted entanglement.
The qubits being sufficiently separated still allows individual addressing with
microwave pulses.
In particular, frequency crowding is also avoided, with the nearest unwanted
transition at least \unit{100}{MHz} detuned for all the relevant transitions.
We emphasize that this optimized mechanism is a general principle for quantum
information processing, whereby destructive interference can be engineered on a
multi-qubit state to mutually cancel out level repulsions coming from nearby
levels and thereby enable both driving of local and nonlocal transitions from
the level.

Lastly, the decoherence rate is only marginally increased in the
quasi-dispersive regime relative to the bare rate, cf.~\Fig[c]{field-free}.
At the point marked X in Figs.~\ref{fig:field-free},~\ref{fig:octmap} the
corresponding error limit is shown in \Fig[b]{qsl} (blue dashed versus solid
gray line).  Thus, point X in Figs.~\ref{fig:field-free},~\ref{fig:octmap}
provides an effective optimum for these combined error mechanisms.
The exact corresponding parameters are listed in \Fig[c]{field-free}.

\subsection{Quantum speed limit for a universal set of gates}
\begin{figure*}[tb]
  \centering
  \includegraphics{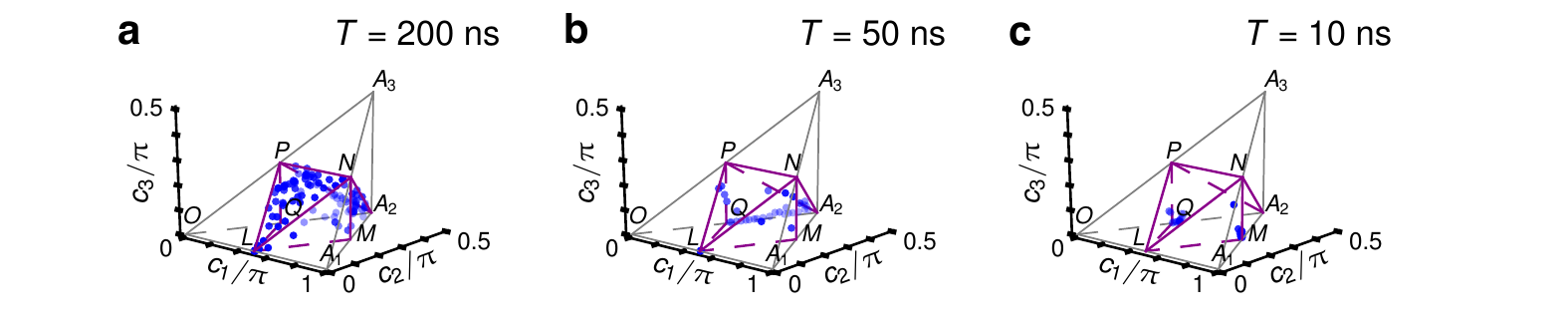}
  \caption{Location in the Weyl chamber of all perfect entanglers (blue dots,
    lighter shades farther back) reached by the maximization of entanglement
    in \protect\Fig[g--i]{octmap}.
   The total Weyl chamber and the polyhedron of perfect entanglers are indicated
   by thin gray and bold magenta lines, respectively.}
\label{fig:weyl}
\end{figure*}
A universal set requires one entangling gate, with many choices available.
We therefore analyze which perfect entanglers can be implemented before
determining the quantum speed limit for a universal set of gates.
To this end, we employ the representation of two-qubit gates in the Weyl chamber
(see Methods). \Figure{weyl} indicates the perfect entanglers that are
successfully implemented in \Fig[g--i]{octmap}.
For long gate durations (\unit{T=200}{ns}, \Fig[a]{weyl}), a large part of the
polyhedron of perfect entanglers is covered.
That is, nearly any perfect entangler can easily be implemented.
This changes as the pulse duration gets shorter.
For \unit{T=50}{ns} (\Fig[b]{weyl}), the reached perfect entanglers are focused
around the L point (diagonal gates), and the lines Q--A$_2$ and Q--P, M--N
(local equivalence classes of $\sqrt{\text{iSWAP}}$--$\sqrt{\text{SWAP}}$).  We
can therefore empirically identify the dominant effective qubit-qubit
interaction term as a linear combination of $\SigmaX\SigmaX$, $\SigmaY\SigmaY$,
and $\SigmaZ\SigmaZ$.
For very short gate durations (\unit{T=10}{ns}, \Fig[c]{weyl}), the obtained
gates cluster strongly around the points Q and M (local equivalence class of
$\sqrt{\text{iSWAP}}$), corresponding to a dominant interaction term
$\SigmaX\SigmaX + \SigmaY\SigmaY = \SigmaPlus\SigmaMinus +
\SigmaMinus\SigmaPlus$.
This is in agreement with the interaction obtained from effective two-qubit
models in the dispersive regime~\cite{BlaisPRA2004}.
Our results suggest that targeting  $\sqrt{\text{iSWAP}}$ (or a gate that is
locally equivalent to $\sqrt{\text{iSWAP}}$) is natural also in the
non-dispersive case.

\begin{figure}[tb]
  \centering
  \includegraphics{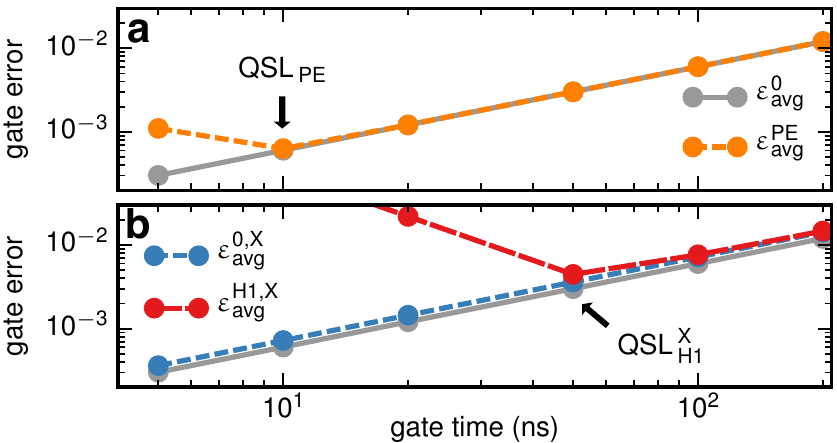}
  \caption{Quantum speed limit (QSL) for the generation of a perfect entangler,
    and a local quantum gate. (a) Minimum gate error $\epsavg^{\PE}$ of perfect
    entanglers over the entire parameter landscape, compared to the
    lifetime-limited error bound $\varepsilon_{\avg}^0$ defined in
    Eq.~\eqref{eq:eps_0}. The minimum in $\epsavg^{\PE}$ represents the quantum
    speed limit for the implementation of a perfect entangler.  (b) Gate error
    $\epsavg^{\text{H1,X}}$ of a Hadamard gate on qubit 1 at point
    X (Figs.~\ref{fig:field-free},~\ref{fig:octmap}) in the optimal QuaDiSQ
    regime, compared to the  lifetime-limited error bound,
    $\epsavg^{0,\text{X}}$, at point X, which differs from the global limit
    $\epsavg^{0}$ by a factor of 1.2 due to the increase in the effective decay
    rate, cf.~\protect\Fig[c]{field-free}. The minimum in
    $\epsavg^{\text{H1,X}}$ may be taken as an estimate of the global quantum
    speed limit for implementing a universal set of gates.
}
\label{fig:qsl}
\end{figure}
Analysis of the gate fidelities of the perfect entanglers allow us to extract
the quantum speed limit for the generation of entanglement.
The smallest error of any perfect entangler in \Fig[g--i]{octmap} is shown as
$\epsavg^{\PE}$ in \Fig[a]{qsl}, together with results for additional gate
durations.
$\epsavg^{\PE}$ is to be compared to the error $\epsavg^{0}$ due to qubit decay,
Eq.~\eqref{eq:eps_0}.
For \unit{T > 10}{ns}, we find the errors to be limited by the qubit lifetime.
For \unit{T = 5}{ns}, the error increases significantly from the
lifetime-limited bound, indicating that the main source of error is now the lack
of capability to create entanglement.
We thus find \unit{T \approx 10}{ns} to be the minimum time necessary to build
entanglement.

Next, we determine the quantum speed limit for \emph{local} gates in the QuaDiSQ
regime, using the parameters of point X. The smallest gate error of a Hadamard
gate on qubit 1 ($\Hadamard \otimes \identity$) is shown as
$\epsavg^{\text{H1},\text{X}}$ in \Fig[b]{qsl}.
We find that the Hadamard gate can be implemented near a lifetime-limited error
bound up to a gate duration of \unit{50}{ns}.
Targeting the remaining single-qubit gates in the universal set, $\identity
\otimes \Hadamard$, $\Phase_{\pi/8} \otimes \identity$, and $\identity \otimes
\Phase_{\pi/8}$ (with $\Phase_{\pi/8} = \diag[1, \exp(-i \pi/4)]$) yields
comparable errors.
We therefore identify \unit{50}{ns} as the speed limit for single-qubit
operations.

For implementing a general perfect entangler at point X with a gate error close
to the theoretical limit, we find a minimum gate duration of \unit{T=20}{ns},
located in the Weyl chamber on the line Q--P. This is slightly larger than the
\emph{global} limit of \unit{10}{ns}.
It is worth noting that this entanglement rate is nonetheless much faster than
what is achievable through resonant qubit-coupling (typically limited to the
inverse of the cavity-mediated qubit-qubit coupling, $\approx \unit{20}{MHz}$ at
point X). This suggests that even for a frequency-tunable architecture, it can
be advantageous to supplement static qubit coupling with microwave drives for
even faster gates.

When decomposing an arbitrary computation of a quantum computer into a universal
set of single and two-qubit gates, it is important to use an entangling
operation in this set that yields the smallest possible number of elementary
gates.
To this end, the perfect entangler must be taken from  a small subset of
``special'' perfect entanglers.
Specifically, gates along the line L--A$_2$ in the Weyl chamber are known to be
efficient~\cite{RezakhaniPRA2004}.
The mid-point of this line, known as the BGATE (see Methods), allows for the
most efficient realization of a universal set.
It requires at most two applications in the decomposition of an arbitrary gate,
compared to three for iSWAP\@.
We therefore target a BGATE at parameter point X. Compared to
iSWAP/$\sqrt{\text{iSWAP}}$, this entails small tradeoffs in the gate duration
and in the complexity of the control pulses.
We find a minimum gate duration for a BGATE of \unit{50}{ns}.

\subsection{Realization of a universal set of gates}

\begin{figure*}[tbp]
  \centering
  \includegraphics{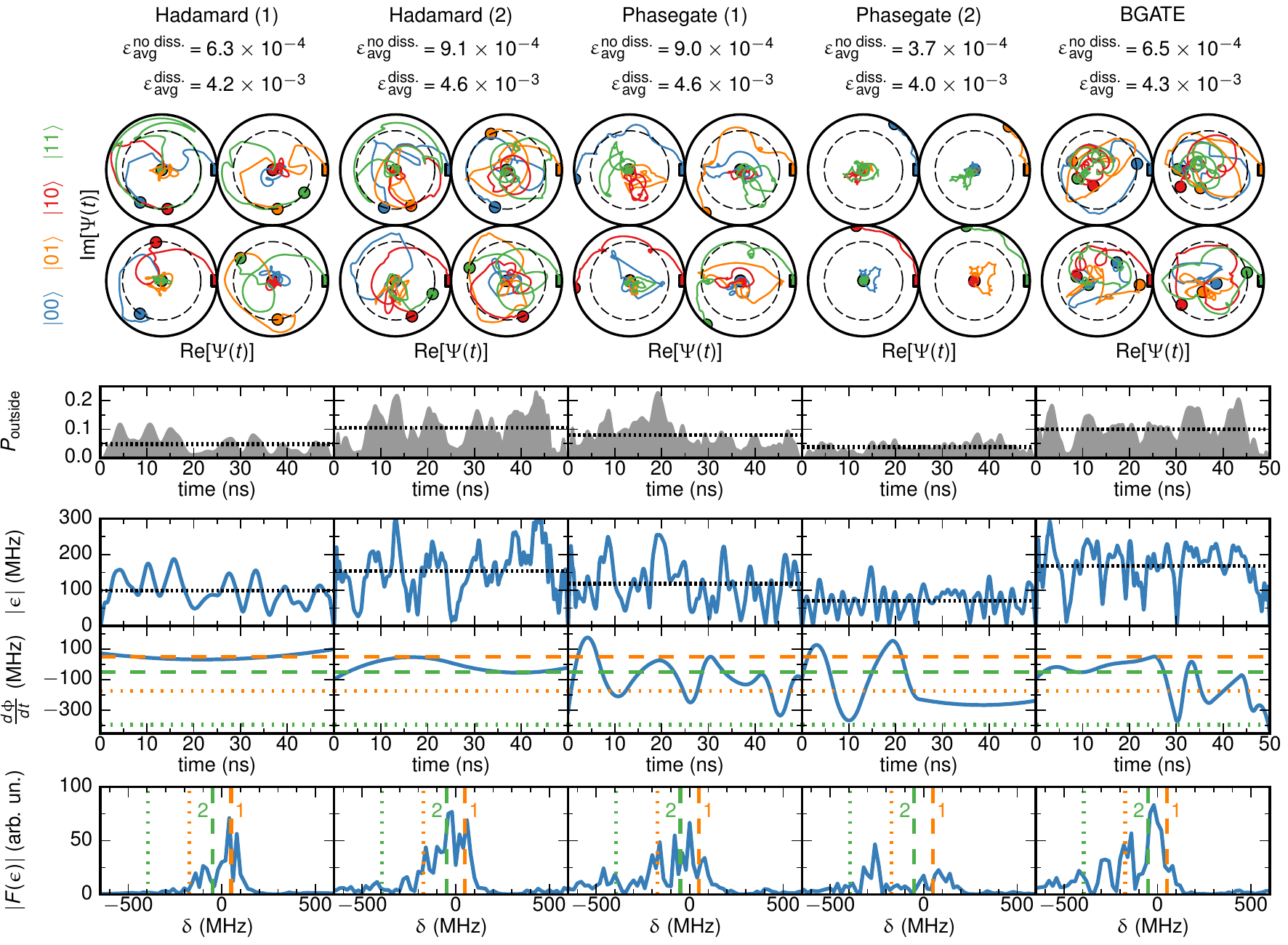}
  \caption{Optimized pulses and dynamics implementing a universal set of
    gates (Hadamard and phase gate $\Phase_{\pi/8}$ on both qubits, and
    non-local BGATE) at the quantum speed limit, for the parameters marked as
    X in Figs.~1,~2.  The gate error $\epsavg^\text{no diss.}$ is evaluated
    without any decay in the system. The dissipative gate error
    $\epsavg^\text{diss.}$ should be compared to the lifetime-limited error
    bound $\epsavg^{0,\text{X}} = 3.6\times 10^{-3}$ for an assumed lifetime of
    \unit{13.3}{\microsec}.  The dynamics are shown for each of the four logical
    basis states, in terms of amplitude and phase of the projection onto the
    logical basis states (color-coded) within the unit circle.  The dashed
    circle marks an  amplitude of $1/\sqrt{2}$.  The values at initial (final)
    time are indicated by the colored squares (bullets).  The population $\Pout$
    outside of the logical subspace is plotted over time.  The properties of the
    optimized pulse $\epsilon(t)$ for each gate are shown, from top to bottom,
    in terms of the pulse amplitude $\Abs{\epsilon}$, the (smoothed) derivative
    of the complex phase $d\phi/dt$, and the spectrum $\Abs{F(\epsilon(t))}$
    for frequencies $\delta$ relative to the rotating frame at $\omega_r/2\pi
    = \unit{5.9325}{GHz}$.  The derivative $d\phi/dt$ gives an approximation for
    $\delta(t)$.  The dressed qubit frequencies are indicated as green and
    orange dashed lines, the dressed ``anharmonic transitions'' $\Ket{1}
    \rightarrow \Ket{2}$ as green and orange dotted lines.  The black dotted
    lines correspond to the mean value of $\Pout$ and $\Abs{\epsilon}$,
    respectively.
    }
\label{fig:universal-pulses}
\end{figure*}

The optimized pulses and the corresponding dynamics (in the interaction picture)
for the entire universal set of gates, consisting of $\Hadamard \otimes
\identity$, $\identity \otimes \Hadamard$, $\Phase_{\pi/8} \otimes \identity$,
$\identity \otimes \Phase_{\pi/8}$, and BGATE, are shown in
\Fig{universal-pulses}.
The average gate error over the entire universal set is $4.3 \times 10^{-3}$,
compared to the lifetime-limited error bound of $\epsavg^{0} = 3.6 \times
10^{-3}$.
In all cases, the achieved gate error is within a factor of 1.3 of the
lifetime-limited error bound.
It is correlated with the mean of the population outside the logical subspace,
$\Pout$, which in turn is correlated with the mean pulse amplitude (black doted
horizontal lines).  Population that is excited to these higher levels is more
strongly affected by dissipation, because the decay rate scales as the square
root of the (bare) quantum numbers.
In fact, evaluating the gate error without dissipation ($\epsavg^{\text{no diss.
}}$ in \Fig{universal-pulses}) shows that the error is dominated by the decay of
the qubit, not by failure to implement the desired gate.
As qubit lifetimes increase with further technological advances of the transmon
platform, errors will approach the value $\epsavg^{\text{no diss.
}}$, consistently below the quantum error correction limit.

For all gates, we obtained the lowest error when the optimization was performed
in a rotating frame with frequency centered exactly between the two qubit
frequencies.
While for the implementation of the Hadamard gate, the spectra show active
frequencies largely around the two qubit transitions (indicated by the dashed
green and orange lines), for the phase gate as well as the BGATE, the pulses
also have strong off-resonant components.
These are predominantly to the left of the qubit transitions, and thus affect
the anharmonic transitions more strongly, driving pathways outside of the
logical subspace.

The derivative of the complex phase, $d\phi/dt$ in \Fig{universal-pulses}
provides a rough estimate of the time-frequency characteristics of the optimized
pulse.
For several of the pulses, we find distinct differences in the active
frequencies over the duration of the pulse.
For example, for $\identity \otimes \Hadamard$, the pulse alternates between the
two qubit frequencies.
For $\identity \otimes \Phase_{\pi/8}$, the strongly off-resonant driving is
interrupted by two brief periods of near-resonant driving.
For the BGATE, the first half of the pulse acts near-resonant on the two qubits,
whereas in the second half, strong off-resonant kicks are applied.
It will be interesting to see in future work whether the observed feature of
alternating periods of near-resonant and off-resonant driving may be exploited
in an \emph{analytic} design of control fields near the quantum speed limit.

In practice, any waveform generator will also have bandwidth and filtering
restrictions that must be taken into
account~\cite{MotzoiPRA2011,JaegerPRA14,TheisPRA2016}.
In order to identify the quantum speed limit, we have not considered any such
restrictions.
Consequently, the pulses have bandwidths in the 500--\unit{1000}{MHz} range.
While this is well within reach of the most current waveform
generators~\cite{1703.00942}, it may be beyond the limitations of older devices
typically used in experiments~\cite{MotzoiPRA2011}.
In such a case, moving away from the speed limit of $\unit{T=50}{ns}$ to e.g.
$\unit{T=100}{ns}$ may provide several practical advantages, as the longer gate
durations give room for applying spectral constraints.
Furthermore, gate errors will approach the lifetime limit more closely, giving
some room to compensate for imperfections in the control scheme.
In the supplementary material, we show pulses implementing a universal set of
gates at $\unit{T=100}{ns}$ with a spectral constraint of $\unit{\pm 200}{MHz}$
around the center of the rotating frame.

Robustness with respect to fluctuations can also be added as an explicit
optimization objective~\cite{GoerzPRA2014}.
For example, for a Hadamard gate on transmon 1 at a longer gate duration of
$\unit{T=100}{ns}$, we find that by simultaneously optimizing over multiple
variations of the system we can account for errors in the pulse amplitude of 1\%
with relative ease, incurring a worst-case drop in the gate error from
$7.3\times 10^{-3}$ to $7.6 \times 10^{-3}$, or $8.9 \times 10^{-5}$ to $3.7
\times 10^{-4}$ without spontaneous decay of the qubit.
In general, both technical constraints and noise sources should be addressed
with respect to a specific experimental setup.
Optimal control techniques towards this end are readily
available~\cite{DongSR2015, AllenPRA2017}.

\section{Discussion}

Superconducting qubits with a shared transmission line come with great
tunability in system parameters.
We have addressed the question of how to choose these parameters in order to
implement a universal set of gates with the best possible errors and shortest
possible gate durations.
We have found that the parameter landscape may be fully characterized by the
qubit-qubit detuning in units of the anharmonicity and the qubit-cavity detuning
in units of the qubit-cavity coupling.
Analysis of the field-free qubit dynamics revealed novel strategies for
implementing both entangling and local quantum gates.
Resonances between qubit levels, or with the cavity, can generate very large
interactions even without any external drive.
In these regions of the parameter space, we showed that a quantum speed limit of
\unit{10}{ns} can be obtained for the controlled generation or removal of
entanglement.
However, a strong, always-on interaction cannot yield a fast universal set of
quantum gates with low error, due to the difficulty to generate specific single
qubit gates.
These regions of the parameter space are therefore only  of interest to setups
that employ tunable qubits.
For fixed-frequency qubits, high-fidelity quantum gates are best implemented
with system parameters that do not yield any static interaction but where strong
interaction can be engineered in a time-dependent fashion by a suitably shaped
microwave pulse.

A key result of this work is the demonstration that the conditions for the
realization of a universal set of gate for short gate durations are best met
outside of the dispersive regime.
Parameter regimes where the dispersive approximation is not valid have remained
under-explored to date, since the Hamiltonian in this regime cannot easily be
approximated to an analytically treatable model.
In this work, we have used a fully numerical approach to explore the complete
parameter space without restrictions and thereby identified a novel parameter
regime as the optimal operating regime for universal quantum computing in which
the global quantum speed limits for a given architecture are attained.
In this new regime, which we term the \emph{Quasi-Dispersive Straddling Qutrits}
(QuaDiSQ) regime, we have shown that two critical but competing requirements for
realizing a universal set of gates are achieved through destructive interference
of multiple near-resonances.
Specifically, one can maintain the near-resonances that allow for fast
entangling gates, while also minimizing static qubit-qubit interaction allowing
for the implementation of local gates.

Using a universal set of gates consisting of the non-local BGATE, as well as
Hadamard and phase gates on each of the two qubits, we derived control protocols
to realize gates with errors within a factor of 1.3 of the lifetime-limited
error bound for a gate duration of \unit{50}{ns}.
Given this identification of the new QuaDiSQ regime and characterization of the
fundamental quantum speed limits, further requirements and constraints of a
specific experimental setup can be taken into account.
For example, by extending the gate duration to \unit{100}{ns}, we can enforce a
spectral width of the pulses of $\unit{\pm 200}{MHz}$ around the center
frequency.
Robustness to experimental parameters such as variations in the pulse amplitude
can further be taken into account for specific setups.

Other choices for the entangling operation are conceivable as well.
For long gate durations, almost all perfect entanglers can be implemented.
When the gate duration is shortened, only the $\sqrt{\text{iSWAP}}$ gate plus
all gates that differ from $\sqrt{\text{iSWAP}}$ by local operations survive.
This indicates that $\sqrt{\text{iSWAP}}$ is the most natural  entangling gate
for the transmon architecture.
This is in agreement with earlier findings obtained  in the dispersive
regime~\cite{BlaisPRA2004}, suggesting that  effective two-qubit models may be
still be qualitatively correct even when they break down quantitatively.

The approach advocated here of identifying a few key parameters and exploring
those with quantum optimal control is not limited to circuit QED platforms nor
quantum computation.
In fact, the paradigm of quantum technology is to engineer an often complex
quantum system to resemble a comparatively simple model Hamiltonian (an
anharmonic ladder system in the case of the transmon).  Quantum engineering
ensures, on one hand, that only a few parameters will be relevant in the
analysis of the possible dynamics.
On the other hand, isolation of the required quantum features typically implies
limited control over the system.
Ensuring the desired functionality (here a universal set of quantum gates)
despite limited control (no or limited tunability of the qubits in the present
case) is possible by harnessing some of the complexity of the quantum system
(using multiple interfering sources of entanglement, as in the QuaDiSQ regime in
our example). This paradigm finds many realizations at the interface of quantum
optics and solid-state physics, for the design of devices to tackle tasks such
as sensing of external fields with quantum-enhanced sensitivity or transmission
of quantum information with cryptographic capability.
Quantum optimal control is essential to achieving this goal.

\section{Methods}

\subsection{Optimization procedure}

The results shown in \Fig{octmap} are the result of optimal control theory
applied to a large sample of parameter points $(\Delta_2/\alpha, \Delta_c/g)$.
At each point, a multi-stage optimization procedure~\cite{GoerzEPJQT2015} is
employed to either minimize or maximize entanglement, proceeding in three
stages.

In the first stage, we presume the control pulse to take the form
\begin{equation}
  \epsilon(t) = E_0 B(t) \cos(\omega_{r} t)
  \stackrel{\text{RWA}}{\longrightarrow}
  E_0 B(t) \,,
  \label{eq:pulse}
\end{equation}
where $B(t)$ is a Blackman shape, similar to a Gaussian, but exactly zero at
initial and final time.
In the rotating frame, the driving frequency $\omega_r$ vanishes and instead is
reflected in the Hamiltonian~\eqref{eq:ham_full}.
The width of the shape is constant, extending over the entire duration.
Thus, the control problem has  only two free parameters, the peak amplitude
$E_0$ and the driving frequency $\omega_{r}$.
Note that a complex $\epsilon(t)$ would be equivalent to a time-dependent phase
of the pulse in the non-rotating frame.
However, we find that the results in \Fig{octmap} do not change significantly
when allowing for multiple frequency components or allowing a non-zero phase.

The first optimization stage consists in choosing  $\omega_{r}$  randomly from
within a range of \unit{1}{GHz} around the qubit and cavity frequencies, and
scanning the pulse amplitude $E_0$ systematically for values $\unit{\in
[10,900]}{MHz}$.
The best pulses are selected according to the functionals
\begin{align}
  \label{eq:J_PE_simplex}
  J^{\simplex}_{\PE} &= 1 - C (1-\varepsilon^{\min}_{\pop}) \,,\\
  J^{\simplex}_{\SQ} &= 1 - (1-C) (1-\varepsilon^{\min}_{\pop})\,. \label{eq:J_QS_simplex}
\end{align}
This takes into account both the concurrence $C$ and the error due to population
loss,
\begin{equation*}
  \varepsilon^{\min}_{\pop} = 1-\min_i \Vert \Op{U} \ket{i} \Vert; \quad
  \ket{i} \in \{\ket{00}, \ket{01}, \ket{10}, \ket{11}\}\,,
\end{equation*}
where $\Op{U}$ is the projection of the time evolution operator onto the logical
subspace.
The functional is written as a product, since the concurrence is only
well-defined for a population loss error near zero.

The selected pulses are the starting point for the second stage, a simplex
(Nelder-Mead) optimization of the two free parameters $E_0$ and $\omega_{r}$ in
Eq.~\eqref{eq:pulse}, using the same optimization functionals
\eqref{eq:J_PE_simplex},~\eqref{eq:J_QS_simplex}.
The third and last stage of  optimization relaxes the constraint imposed by the
simple analytical form~\eqref{eq:pulse}, and uses Krotov's
method~\cite{ReichJCP12} to continue optimization of  $\epsilon(t)$ for an
arbitrary perfect entangler, respectively an arbitrary local gate.
As any gradient-based optimization approach, Krotov's method requires a
differentiable functional.
Since the concurrence is non-analytic in the control, we cannot employ
Eqs.~\eqref{eq:J_PE_simplex}, \eqref{eq:J_QS_simplex} and need to resort to
optimization in the Weyl chamber as described below.
The pulse shape is now allowed to take complex values.
The total optimization functional also includes a term $\tr[\Op{U}^\dagger
\Op{U}]/4$ to penalize loss from the logical subspace~\cite{MullerPRA11}.

The optimization for a complete universal set of gates uses a similar
three-stage procedure as outlined above, but optimizes towards the BGATE (up to
local operations, see below) and towards the four local gates using the  gate
overlap with the target gate \Op{O},
\begin{equation*}
  \Fsm = \Abs{\tr\left[ \Op{O}^\dagger \Op{U}\right]}^2;\qquad
  \Jsm = 1 - \Fsm\,.
\end{equation*}

\subsection{Optimization in the Weyl chamber}

Any two-qubit gate $\Op{U} \in \SU(4)$ can be written according to the Cartan
decomposition~\cite{MakhlinQIP2002,ZhangPRA2003},
\begin{equation*}
 \Op{U} = \Op{k}_1 \exp\left[
   \frac{i}{2} \left(%
     c_1 \SigmaX\SigmaX
   + c_2 \SigmaY\SigmaY
   + c_3 \SigmaZ\SigmaZ
    \right)
  \right] \Op{k}_2\,,
\end{equation*}
where $\SigmaX$, $\SigmaY$, $\SigmaZ$ are the Pauli matrices, and $\Op{k}_{1,2}
\in \SU(2) \otimes \SU(2)$ are single-qubit, ``local'' operations.
Taking into account symmetries, the coefficients $c_1$, $c_2$, $c_3$ take values
$c_1 \in [0, \pi]$ and $c_2, c_3 \in [0, \pi/2]$.
They may be interpreted as coordinates in a three-dimensional space; all the
points that represent the possible two-qubit gates up to local operations then
form a quarter-pyramid called the Weyl-chamber.
It is depicted in \Fig{weyl}.
The named edges and vertex midpoints of the Weyl chamber correspond to some of
the ``standard'' two-qubit gates (CNOT, CPHASE, SWAP, iSWAP, etc.
). The point $(\pi/2, \pi/4, 0)$, i.e., the midpoint of the line L--A$_2$,
defines the BGATE that is the optimal perfect entangler for the universal set of
gates, with the canonical representation
\begin{equation*}
  \text{BGATE} =
  \begin{pmatrix}
    \cos\frac{\pi}{8} & 0 & 0  & i \sin\frac{\pi}{8} \\
    0 & \cos\frac{3\pi}{8} & i \sin\frac{3\pi}{8} & 0 \\
    0 & i \sin\frac{3\pi}{8} & \cos\frac{3\pi}{8} & 0 \\
    i \sin\frac{\pi}{8} & 0 & 0 & \cos\frac{\pi}{8}
  \end{pmatrix}\,.
\end{equation*}

Separating a quantum gate into local and non-local contributions through the
Cartan decomposition allows to optimize for a quantum gate \emph{up to local
operations} by minimizing the ``local invariants functional'' $\JLI$. It % that
evaluates the geometric distance to a target point in the Weyl
chamber~\cite{MullerPRA11}.
This is especially appropriate in the context of universal quantum gates, as we
assume that arbitrary single-qubit gates can be implemented.

The majority of two-qubit entanglers which form a polyhedron with the edge
points L, M, A$_2$, Q, P, and N, indicated in magenta in \Fig{weyl}.
In order to optimize for an arbitrary perfect entangler using a gradient-based
method, one can define a ``perfect entanglers functional'' $\JPE$ that minimizes
the geometric distance to the surface of the polyhedron~\cite{WattsPRA2015}.
Note that the opposite, optimizing for an arbitrary local gate (non-entangling),
is achieved by applying the local invariants functional towards the points O or
A$_1$.
Only when we need to implement a \emph{specific} gate (e.g., the single-qubit
gates in the universal set) do we employ the gate overlap $J_\mathrm{sm}$.

Any of the functionals defined in the Weyl chamber assume that $\Op{U} \in
\SU(4)$.
When the model includes levels outside the logical subspace, as is the case
here, some of the population may remain in those spurious levels at final time
$T$.
The resulting projection $\tildeOp{U}$ to the logical subspace may no longer be
unitary.
In this case, we must include the loss in the functional~\cite{MullerPRA11}, and
evaluate $\JLI$ or $\JPE$ using the \emph{closest unitary} gate $\Op{U}$.
If $\tildeOp{U}$ has the singular value decomposition $\tildeOp{U} = \Op{V} \,
\Op{\Sigma} \, \Op{W}^{\dagger}$, then the closest unitary is
\begin{equation*}
  \Op{U} = \arg \min_{\Op{U}'} \Norm{\tildeOp{U} - \Op{U}'}
          = \Op{V} \, \Op{W}^\dagger\,.
\end{equation*}

When evaluating the optimization success using $\JLI$, we must find the local
operations $\Op{k}_1$, $\Op{k}_2$  in the Cartan decomposition such that
$\epsavg$ is minimized.
This is done by parametrizing an arbitrary single-qubit gate as
\begin{equation*}
  \Op{U}_{1q} = e^{i \phi} \begin{pmatrix}
          \cos\theta \, e^{i\phi_1}   & \sin\theta \, e^{i\phi_2}\\
         -\sin\theta \, e^{-i\phi_2}  & \cos\theta \, e^{-i\phi_1}\\
        \end{pmatrix}\,.
\end{equation*}
With $\Op{k}_1$ and $\Op{k}_2$ consisting of two single-qubit gates each, this
gives a total of 16 free parameters that are easily determined through numerical
minimization.
For $\JPE$, the procedure is the same, except that the appropriate value of the
Weyl coordinates $c_1, c_2, c_3$ must first be determined by projecting $\Op{U}$
onto the surface of the polyhedron of perfect entanglers (assuming $\Op{U}$
itself is not already a perfect entangler).

\subsection{Average Gate Error}

While the functionals $\JLI$, $\JPE$, or $\Jsm$  are suitable for numerical
purposes to steer an optimization, they are not directly accessible to
measurement and can thus not be used to objectively evaluate the success of gate
implementation.
The experimentally relevant measure of success for implementing the target gate
\Op{O} with a dynamical map $\DynMap$ is the \emph{average gate fidelity}
\begin{equation*}
  F_{\avg} = \int \langle \Psi | \Op{O}^{\dagger}
              \DynMap(T,0)[\ket{\Psi}\!\bra{\Psi}]
              \Op{O} | \Psi \rangle \, d\Psi\,,
\end{equation*}
respectively the average gate error $\epsavg = 1 - F_{\avg}$.
The error is easily evaluated numerically~\cite{PedersenPLA07}.
We report all gate errors in terms of $\epsavg$, independent of the functional
used in the optimization.

In order to derive a lower limit for the achievable gate error for a given qubit
decay rate, i.e., the lifetime-limited error bound, we consider the Hamiltonian
of Eq.~\eqref{eq:ham_full} with $g=0$ and $\epsilon(t) \equiv 0$.
The cavity can then be integrated out, and the transmon Hilbert space can be
truncated to two levels.
In the interaction picture and without dissipation, there is no time evolution
($\Op{O} = \identity$). The Liouville-von-Neumann equation can then be solved
analytically for the qubit decay rate $\gamma$.
Plugging the result into the formula for $\epsavg$ yields
\begin{equation*}
  \epsavg^{0}(\gamma,T)
  = \frac{3}{4} - \frac{3 \, e^{- \gamma T}}{10}
    - \frac{e^{- 2 \gamma T}}{20}
    - \frac{e^{- \frac{\gamma T}{2}}}{5}
    - \frac{e^{- \frac{3\gamma}{2} T}}{5}\,.
\end{equation*}
For $\gamma T \ll 1$, this can be linearized to Eq.~\eqref{eq:eps_0}.

\subsection{Equation of motion}

In order to evaluate the average gate error for any of the optimized pulses, we
numerically solve the full Liouville-von-Neumann equation
\begin{equation}
  \frac{\partial}{\partial t} \Op{\rho}
  = -i [\Op{H}, \Op{\rho}] + \sum_{i=1}^{3} \left(
      \Op{A}_i \Op{\rho} \Op{A}_i^{\dagger}
      - \frac{1}{2} \left\{ \Op{A}_i^{\dagger} \Op{A}_i, \Op{\rho} \right\}
    \right)\,,
  \label{eq:liouville}
\end{equation}
with the Hamiltonian~\eqref{eq:ham_full}, and the Lindblad operators $\Op{A}_1 =
\sqrt{\gamma} \Op{b}_1$, $\Op{A}_2 = \sqrt{\gamma} \Op{b}_2$, and $\Op{A}_3 =
\sqrt{\kappa} \Op{a}$.
The Hilbert space of the two transmons and the cavity are truncated at 5 and 6
levels, respectively.
This has been checked to be sufficient for all pulses considered here.

For the purpose of optimization, solving Eq.~\eqref{eq:liouville} is numerically
too expensive.
Instead, we solve the Schr\"odinger equation with an non-Hermitian effective
Hamiltonian to mimic population loss,
\begin{equation}
\Op{H}_{\text{eff}}  = \Op{H} - \frac{i\hbar}{2} \sum_i \Op{A}_i^{\dagger} \Op{A}_i\,.
\end{equation}
We also increase the decay rate of the highest qubit and cavity level to
infinity as a method to ensure that the optimized pulses are well-described in
the truncated Hilbert space.
We stress that this non-Hermitian Hamiltonian is only used to effectively
penalize population in strongly dissipative levels during the optimization.
The final optimized pulses are then evaluated by solving
Eq.~\eqref{eq:liouville}; all reported dynamics and errors are obtained from
this density matrix evolution.
Both the Liouville-von-Neumann equation and the Schr\"odinger equation with a
non-Hermitian Hamiltonian can be solved efficiently and to high precision using
an expansion into Newton polynomials in a Krylov
subspace~\cite{Tal-EzerSJSC2007}, as implemented in the Fortran QDYN package.

\subsection{Acknowledgements}

\begin{acknowledgments}
We thank the Kavli Institute for Theoretical Physics for their hospitality and
for supporting this research in part by the National Science Foundation Grant
No.\ PHY11-25915.
Financial support by the DAAD under Grant No.\ PPP USA 54367416 and the National
Science Foundation under the Catalyzing International Collaborations program
(Grant No.\ OISE-1158954) is gratefully acknowledged.
CPK was also supported by the DFG project ``Control of open systems'', Grant
No.\ KO 2301/11-1.
Publication made possible in part by support from the Berkeley Research Impact
Initiative (BRII) sponsored by the UC Berkeley Library.
\end{acknowledgments}

\subsection{Competing interests}

The authors declare no competing interests.

\subsection{Author contributions}

M.\,H.\,G.\ performed the numerical calculations and data analysis.
All authors contributed to the planning, discussion of results, and preparation
of the manuscript.

\subsection{Data availability}

The numerical data that support the findings of this study are available from
the corresponding author upon request.

\subsection{Code availability}

The QDYN library for quantum dynamics and control in Fortran was used for all
calculations of the system dynamics and optimization using Krotov's method.
It is available upon request through \url{https://www.qdyn-library.net}.
The Nelder-Mead simplex optimizations were performed using the SciPy Python
library (v0.17), freely available at \url{https://scipy.org}.

~\clearpage
\pagestyle{empty}


\section*{Supplementary material}

# Supplementary Material:
# Charting the circuit QED design landscape using optimal control theory

Michael H. Goerz,[1, *] Felix Motzoi,[2, 3, †] K. Birgitta Whaley,[2, 3] and Christiane P. Koch[1]

[1] *Theoretische Physik, Universität Kassel, Heinrich-Plett-Str. 40, D-34132 Kassel, Germany*
[2] *Berkeley Center for Quantum Information and Computation, Berkeley, California 94720 USA*
[3] *Department of Chemistry, University of California, Berkeley, California 94720 USA*

Supplementary Figure 1 shows the optimized pulses and dynamics for a universal gate set at $T = 100$ ns, with a spectral constraint of $\pm 200$ MHz around the central frequency of the rotating frame. The resulting pulse spectra are sufficiently simple to be within reach of current pulse shaping technology. Numerically, the constraint was enforced by multiplying the optimized pulses with a filter function after each iteration of Krotov's method.

The results compare to Fig. 5 in the main text, which shows and unconstrained set of pulses at the quantum speed limit of $T = 50$ ns. While for $T = 50$ ns, the lowest gate errors were obtained when all pulses are set in a rotating frame centered exactly between the two (dressed) qubit frequencies, here we find that the this is only optimal for the entangling BGATE. All of the local gates are most successful with a spectrum centered around the dressed frequency of transmon 2. We take this as a consequence of the narrower pulse spectrum in conjunction with the qubits being in the "straddling" (QuaDisQ) regime. As the frequency of transmon 2 is halfway between the first and second transition of qubit 1 (orange dashed and dotted lines), this gives more room for the anharmonic transition to take part in the gate realization.

The gates errors reach closer to the lifetime limit than the errors obtained for shorter pulses; for the Hadamard gate on either transmon, we reach the limit of $7.2 \times 10^{-3}$ almost exactly. If the lifetime of the transmon can be increased through further technological advances, gate errors may reach an order as low as $10^{-5}$.

---

[*] Current affiliations: Edward L. Ginzton Laboratory, Stanford University, Stanford, CA 94305, USA., and U.S. Army Research Laboratory, Computational and Information Sciences Directorate, Adelphi, MD 20783, USA.
[†] Current affiliation: Department of Physics and Astronomy, Aarhus University, 8000 Aarhus C, Denmark.

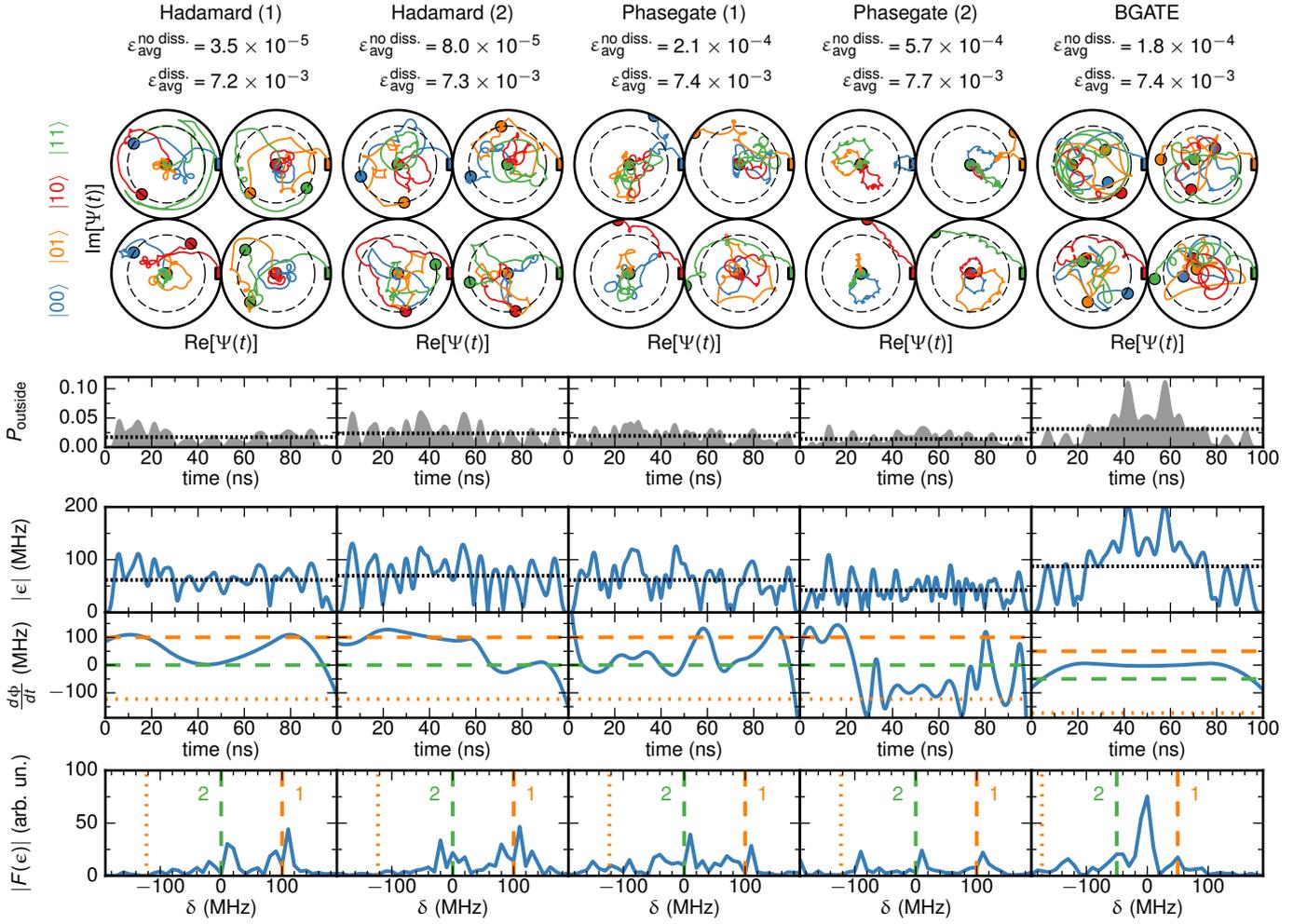

Supplementary Figure 1. Pulses and dynamics implementing a universal set of gates, for a gate duration of 100 ns and a constraint on the spectrum of the optimized pulses of $|\delta| \leq 200$ MHz. The figure compares to the unconstrained optimization for $T = 50$ ns in Fig. 5 of the main text. The dissipative gate errors $\varepsilon_{\text{avg}}^{\text{diss}}$ approach the lifetime-limited $\varepsilon_{\text{avg}}^{0,\text{B}} = 7.2 \times 10^{-3}$, cf. Fig. 4 (b) in the main text.


\end{document}